\documentclass[10pt,aps,pre,twocolumn,notitlepage,superscriptaddress,preprintnumbers]{revtex4-1}
\usepackage[utf8]{inputenc}
\usepackage[utf8]{inputenc} 
\usepackage[T1]{fontenc}

\usepackage{amsmath,amssymb,amsfonts}
\usepackage{bm}
\usepackage{graphicx,color}
\usepackage{verbatim}
\usepackage{float}
\usepackage[caption = false]{subfig}
\usepackage{dcolumn}
\usepackage{natbib}
\usepackage{hyperref}
\usepackage{tikz}
\usetikzlibrary{hobby}
\usepackage{appendix}
\usepackage{ulem}

\newcommand{\RN}[1]{%
	\textup{\uppercase\expandafter{\romannumeral#1}}%
}

\begin{document}
	
	\title{Heider balance under disordered triadic interactions}
	\author{M. Bagherikalhor}
	\email{mahsa.bagherikalhor@gmail.com}
	\affiliation{Department of Physics, Shahid Beheshti University, G.C., Evin, Tehran 19839, Iran}
	\author{A. Kargaran}
	\affiliation{Department of Physics, Shahid Beheshti University, G.C., Evin, Tehran 19839, Iran}
	\author{A. H. Shirazi}
	\affiliation{Department of Physics, Shahid Beheshti University, G.C., Evin, Tehran 19839, Iran}
	\author{G. R. Jafari}
	\email{g\_jafari@sbu.ac.ir}
	\affiliation{Department of Physics, Shahid Beheshti University, G.C., Evin, Tehran 19839, Iran}
	\affiliation{Institute for Cognitive and Brain Sciences, Shahid Beheshti University, G.C., Evin, Tehran, 19839, Iran}
	
	\date{\today}
	
	\begin{abstract}
	The Heider balance addresses three-body interactions with the 
	assumption that triads are equally important in the dynamics of the 
	network. In many networks, the relations do not have the same strength 
	so, triads are differently weighted. Now, the question is how social 
	networks evolve to reduce the number of unbalanced triangles when they 
	are weighted? Are the results foreseeable based on what we have 
	already learned from the unweighted balance? To find the solution, we 
	consider a fully connected network in which triads are assigned with 
	different random weights. Weights are coming from Gaussian probability 
	distribution with mean $\mu$ and variance $\sigma$. We study this 
	system in two regimes: (\RN{1}) the  ratio of $\frac{\mu}{\sigma} \ge 
	1 $ corresponds to weak disorder (small variance) that triads' weight 
	are approximately the same, (\RN{2}) $\frac{\mu}{\sigma} < 1 $ counts 
	for strong disorder (big variance) and weights are remarkably diverse. 
	Investigating the structural evolution of such a network is our 
	intention. We see disorder plays a key role in determining the 
	critical temperature of the system. Using the mean-field method to 
	present an analytic solution for the system represents that the system 
	undergoes a first-order phase transition. For weak disorder, our 
	simulation results display the system reaches the global minimum as 
	temperature decreases whereas for the second regime, due to the 
	diversity of weights, the system does not manage to reach the global 
	minimum.\\
	\end{abstract}
	\maketitle
	\section{\label{sec:level1}Introduction} 
	Balance theory was first introduced by Heider in 1946 as a concept in 
	social psychology \cite{heider1}\cite{heider2}. Heider dedicated sign 
	positive (negative) to pairwise interactions for friendship (enmity) 
	relations. The theory considers triadic relationships; a group of 
	three persons in which interaction between individuals is friendship 
	or enmity. The sign of the product of links shows whether a triadic 
	relationship is balanced $(+1)$ or unbalanced(frustrated) $(-1)$ 
	states. The balanced states are defined when all persons are friends 
	or two friends have a common enemy(even numbers of negative links). 
	These states are formulated as familiar rules that a friend of a 
	friend will be a friend and an enemy of a friend will be an enemy. The 
	essential idea of balance theory is based on reducing the number of 
	unbalanced(frustrated) triads that cause tension in a network of 
	relations. A network will be balanced if each triad is balanced 
	\cite{heider2}\cite{faust}. Later on, Cartwright and Harary expanded 
	the idea into the graph theory and was termed as structural balance 
	theory \cite{cartwright}. In a social network, the evolution of 
	relations is a remarkable issue and each link changes its sign in a 
	way to minimize tension which is a natural trend of human beings. The 
	network evolution results in two final states of balance, either 
	\textit{heaven} or \textit{bipolar}, however, Davis \cite{davis} by 
	presenting some theorems goes beyond bipolar and deal with the 
	clustering of incomplete signed graphs into multiple cliques.
	
	Structural balance theory has a vast application and comprehensive 
	literature in different branches of science including social 
	psychology \cite{heider1,heider2,cartwright}, mathematical sociology 
	\cite{leik}, ecology \cite{saiz}, and studies of international 
	networks \cite{hart,galam,bramson}. A large number of works 
 	in physics have looked structural balance through different lenses \cite{singh,altafini,oloomi,amir,sheykhali,hedayatifar,hassanibesheli,saeedian,fereshteh,razieh}. Belaza \textit{et al.} proposed a generic Hamiltonian and applied statistical physics to investigate balance theory in political networks \cite{belaza1} then, made an extension of signed networks by introducing inactive links \cite{belaza2}.  
 	
 	Many studies address how a network undergoes a dynamic transition from an initial state to \textit{heaven} or \textit{bipolar}. Antal \textit{et al.} studied different dynamic rules for achieving balance. They solved discrete-time dynamics to investigate how an initially 
 	unbalanced society achieves balanced \cite{antal1,antal2}. Kułakowski 
 	\textit{et al.} investigated the continuous-time evolution of social 
 	relations by considering real values for interaction strengths 
 	\cite{kulakowski}. Thereafter,	Marvel \textit{et al.} also analyzed 
 	a continuous-time dynamical system and concluded that the initial 
 	amount of friendliness determines the destination of a network whether 
 	it reaches a global harmony or splits into two clusters with internal 
 	friendly and external hostile relationships \cite{marvel2}. Defining 
 	an energy landscape, addressing the concept of local minimums 
 	so-called jammed states, and their structural dependence on the size 
 	of the network is of concern to Ref. \cite{marvel1}. 
 
 	Investigating the meaning of balance in real-world data has attracted 
 	some attention. Leskovec \textit{et al.} analyzed online signed social 
 	networks using two different theories. Their results showed how these 
 	networks are unbalanced in contrast with the expected view of fully 
 	balanced \cite{leskovec}. Different measures of balance in a signed 
 	network are studied in Ref. \cite{kirkely}. Facchetti \textit{et al.} 
 	computed the global level of balance and confirmed that currently 
 	available social networks are extremely balanced \cite{facchetti}. In 
 	contrast, Ref. \cite{estrada1,estrada2} based on an introduced method 
 	to quantify the degree of balance of any social network, show that 
 	online social networks are in general very poorly balanced. Although 
 	most publications address unweighted social networks to study, several 
 	works try to provide a more complete view of networks by considering 
 	heterogeneity in the intensity of the interactions between individuals 
 	\cite{newman4,newman5,yook,noh,barrat1,barrat2}. It is 
 	worth noting that the pairwise interactions were the center of 
 	attention for years and many physics phenomena are explained based on 
 	them, while triadic interactions as a specific structure of higher 
 	order interactions have been noticed by researchers as well. 
 	Topological aspects of higher-order connectivity have been considered 
 	in Ref. \cite{Tadic1}. This work addresses the expansion of a network 
 	by simplexes of different sizes. Also, in Ref. \cite{Tadic2} the 
 	authors by considering the geometry and dynamics of complex networks 
 	focus on triangle-based interactions. Three-body interactions are the 
 	first step beyond pairwise interactions that managed to find 
 	applications \cite{abbas,zahra}. Besides all efforts 
 	which are done in studying social network, our mind is still strongly 
 	obsessed with the question of what if we consider higher order 
 	interactions and assign different weights to triadic relations. Is it 
 	plausible to study the statistical properties of such a network 
 	theoretically by considering a modified version of structural balance?
 	
	Here, we present a new Hamiltonian and study the triadic interaction 
	dynamics in non zero temperature when triangles are randomly weighted 
	and we introduce weights as a disorder. This model is considering the 
	intensity of friendship and enmity relationships since all communities 
	of relations do not have equal priority.
	 In a network of interactions, people tend to reduce tension in their 
	 relationships so, a relation with less tension is 
	 more popular. We consider temperature as randomness in the social 
	 network under which individuals modify their relationships to resolve 
	 the tension. Links that belong to triangles with larger weights need 
	 more energy to change their signs. We use exponential random graphs 
	 \cite{holland,besag,frank,strauss,wasserman,anderson,snijders1,robins,cranmer,snijders2}
	  as our statistical approach to analyze the introduced Hamiltonian. 
	 We find the expected value of links, weighted two-stars (two-links 
	 connected to a common node), and the energy. In providing a solution 
	 for our Hamiltonian we use the mean-field approach 
	 \cite{newman1,newman2,newman3}.  At last, we compare our results for 
	 calculated statistical variables with simulation; we see the 
	 agreement between theory and simulation is well when we choose 
	 variance to be small in the probability distribution of weights.
	\section{Model}\label{model}
	A concept for the energy of a social network proposed by Marvel 
	\textit{et al.} \cite{marvel1}. They presented a potential energy that 
	is proportional 
	to the sum of all triangles of the network. Subsequently, the 
	Hamiltonian of the structural balance is
	\begin{equation}
	\mathcal{H}(G)=  -{\sum_{i>j>k} S_{ij} S_{jk} S_{ki}},
	\end{equation}
	here, we consider a modified version of structural balance which takes 
	into account random weights for the triads. Weights are quenched 
	meaning that they are constant on the time scale over which the links 
	fluctuate. We could consider the case in which links are weighted 
	however, we assign a value to each triad and let links to be signed. 
	The proposed Hamiltonian of our model is
	\begin{equation}\label{eq1}
    \mathcal{H}(G)= - {\sum_{i>j>k}J_{ijk} t_{ijk}(G)} = -{\sum_{i>j>k}J_{ijk} S_{ij} S_{jk} S_{ki}},
	\end{equation}
	where \textit{$\mathcal{H}(G)$} is the graph Hamiltonian, $\left\{t_{ijk}\right\}$ is the complete set of triads in the network, and $S_{ij}=S_{ji}$ is an element of the adjacency matrix \textbf{S} which its value can be $\left\{\pm{1}\right\}$. Each element is a link exists between nodes $i$ and $j$ and represents the kind of relationship (friendship or enmity). The conjugate field (so-called weight), $J_{ijk}$, for the triad formed on nodes $i$, $j$, $k$ is a real value coming from a Gaussian probability distribution with a specific mean $(\mu)$ and variance $(\sigma)$ 
	\begin{equation}
	P(J_{ijk})=\frac{1}{\sqrt{(2\pi 
			\sigma^2)}}\exp(\frac{-(J_{ijk}-\mu)^2}{2\sigma^2}).
	\end{equation}
	
\begin{figure}[ht]
	\centering
	\includegraphics[scale = .8]{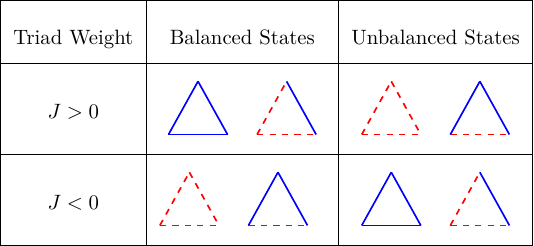}
	\caption{Solid line ( dashed line ) represents positive ( negative ) 
	relationship. $J>0$ : $\{+,+,+\}$ , $\{+,-,-\}$ are balanced, and 
	$\{-,-,-\}$ , 
	$\{+,+,-\}$ are unbalanced states. $J<0$ : $\{-,-,-\}$ , $\{+,+,-\}$ 
	are balance, and $\{+,+,+\}$ , $\{+,-,-\}$ are unbalanced states. A 
	triangle is called balanced if the product of its weight and links 
	signs, $J_{ijk}S_{ij}S_{jk}S_{ki}$, is positive and unbalanced if the 
	product is negative.}
	\label{fig:fig1}
	\end{figure}

	In the case of a uniform field with all weights equal to one, Eq. 
	(\ref{eq1}) reduces to structural balance Hamiltonian which its 
	dynamic in nonzero temperature is studied in Ref. \cite{fereshteh}. 
	In the structural balance, there are two types of 
	triads, balanced and unbalanced, and the network reaches its global 
	minimum when all triads are balanced and the energy equals $-1$. The 
	local minimum occurs where the unbalanced triads exist in the network 
	and updating a link results in energy increment. To investigate the 
	energy landscape in the proposed model, let consider 
	two possible situations of weights $J>0$, $J<0$. Based on 
	Fig.~\ref{fig:fig1} in each case balanced  and 
	unbalanced states are defined independently. According to the 
	Hamiltonian Eq. (\ref{eq1}) balanced states correspond to the positive 
	product of weight and links of a triad $J_{ijk}S_{ij}S_{jk}S_{ki}$, 
	which lessen the energy of the network. The product of weight and 
	links of the unbalanced states is negative and they increase the 
	energy of the network.

	 Our model investigates how the network will reach a balanced state as 
	 a result of the competition of triads with various weights in each 
	 temperature. We show that the temperature has a significant effect on 
	 the dynamics of the network. In the proposed model, the minimum of 
	 energy is different from $-1$ in structural balance theory and 
	 depends on the parameters of the probability distribution of weights 
	 ($\mu, \sigma$). The network experiences a balanced state based on 
	 the distinct concept of balance in positive and negative weighted 
	 triads. Inspired by the exponential random graph, using the Boltzmann 
	 distribution gives us the probability of selecting a specific graph 
	 configuration at a certain temperature within the set of graphs. The 
	 Boltzmann probability in canonical ensemble is, 
	 $\mathcal{P}(G)\propto e^{-\beta \mathcal{H}(G)} $, where $\beta=1/T$.
	\section{Analysis}\label{analysis}
	\subsection{Mean-field solution}
	Considering a fully connected network, we study the weighted 
	structural balance Hamiltonian Eq. (\ref{eq1}) and present a solution 
	for the model based on the mean-field approach which 
	is exact in the limit of large system size. Let rewrite our 
	Hamiltonian as $\mathcal{H}=\mathcal{H'}+ \mathcal{H}_{ij}$, 
	$\mathcal{H}_{ij}$ includes all terms in the Hamiltonian that contain 
	$S_{ij}$,
	\begin{equation}
	\mathcal{H}_{ij}=- S_{ij}{\sum_{k\neq{i,j}} 
		J_{ijk}S_{jk}S_{ki}},
	\end{equation}
	and $\mathcal{H'}$ is the rest of the Hamiltonian that relates to other links. The mean value $\langle S_{ij}\rangle$ of $S_{ij}$ can be calculated as
	\begin{equation}
	\begin{aligned}
	\langle{S_{ij}}\rangle &= (1)\times P(S_{ij}=1)+ (-1)\times P(S_{ij}=-1)\\ 
	&=\frac{1}{\mathcal{Z}}\sum_{\{S'\}}e^{-\beta\mathcal{H}'}\sum_{S_{ij}={\left\{\pm{1}\right\}}}S_{ij}e^{-\beta
	 \mathcal{H}_{ij}}\\
	&=\frac{\sum_{\{S'\}}e^{-\beta\mathcal{H}'}\left[e^{-\beta\mathcal{H}_{ij}(S_{ij}=+1)}-e^{-\beta\mathcal {H}_{ij}(S_{ij}=-1)}\right]}{\sum_{\{S'\}}e^{-\beta\mathcal{H}'}\left[e^{-\beta\mathcal{H}_{ij}(S_{ij}=+1)}+e^{-\beta\mathcal{H}_{ij}(S_{ij}=-1)}\right]}\\
	&=\frac{\left\langle e^{-\beta\mathcal{H}_{ij}(S_{ij}=+1)}-e^{-\beta\mathcal {H}_{ij}(S_{ij}=-1)}\right\rangle_{\mathcal{Z'}}}{\left\langle e^{-\beta\mathcal {H}_{ij}(S_{ij}=+1)}+e^{-\beta\mathcal{H}_{ij}(S_{ij}=-1)}\right\rangle_{\mathcal{Z'}}},
	\end{aligned}
	\end{equation}
	where $\mathcal{Z} = \sum_{_{\left\{ G \right\}}}\exp{(-\beta\mathcal{H}(G))}$ is the partition function and we define: $\mathcal{Z'} = \sum_{_{\left\{ S' \right\}}}\exp{(-\beta\mathcal{H'})}$. Here, $\langle\cdots\rangle_{\mathcal{Z'}}$ indicates average within ensemble of networks over all links except $S_{ij}$. We can expand each of the exponential terms and applying the mean-field approximation term by term. Approximating all correlations of higher order by the correlation between single pairs of links results in, terms like $\langle{(S_{jk}S_{ki})^m}\rangle$ can be written as $\langle{S_{jk}S_{ki}}\rangle^m$ in the mean-field approximation. Following this approximation and naming $p\equiv\langle{S_{ij}}\rangle$, the mean value of links is
	\begin{equation}\label{eq5}
	p = \tanh(\beta\langle{\sum_{k\neq{i,j}}J_{ijk}S_{jk}S_{ki}}\rangle),
	\end{equation}
	\\
	we call the expression within the expectation symbol in Eq. (\ref{eq5}) the effective field each link feels and name it as \textit{per link field}, $Q$, 
	\begin{equation}\label{eq6}
	 Q\equiv\langle \sum_{k\neq{i,j}}J_{ijk}S_{jk}S_{ki}\rangle. 
	\end{equation}

	By analogy to the steps mentioned above, we rewrite our Hamiltonian as $\mathcal{H}=\mathcal{H'}+ \mathcal{H}_{jk,ki}$ 
	\begin{equation}\label{eq7}
		\begin{aligned}
			\mathcal{H}_{jk,ki}&=-S_{jk}(\sum_{l\neq{i,j,k}}J_{jlk}S_{jl}S_{kl})-S_{ki}(
			\sum_{l\neq{i,j,k}}J_{kil}S_{kl}S_{il})\\
			&-J_{ijk}S_{ij}S_{jk}S_{ki},
		\end{aligned}
	\end{equation}	
	  we replace $\sum_{l\neq{i,j,k}}$ by $\sum_{l\neq{i,j}}$ in the 
	  mean-field approximation. By this estimation we are counting the 
	  specific two-star on nodes $i,j,k$ twice but in averaging over the 
	  all two-stars on a link, $N-2$, this outnumbering is ignorable. This 
	  approximation allows us a convenient use of $Q$ definition. Now, we 
	  set up an equation for calculating the mean value of two-stars 
	  $q\equiv\langle{S_{jk}S_{ki}}\rangle$, 
	\begin{widetext} 
	\begin{equation}\label{meanoftwostars}
		\begin{aligned}
		\langle{S_{jk}S_{ki}}\rangle=\frac{\langle e^{-\beta\mathcal{
				H}_{jk,ki}(S_{jk},S_{ki}=1)}-e^{-\beta\mathcal{
				H}_{jk,ki}(S_{jk}=1,S_{ki}=-1)}-e^{-\beta\mathcal{
				H}_{jk,ki}(S_{jk}=-1,S_{ki}=1)}+e^{-\beta\mathcal{
				H}_{jk,ki}(S_{jk},S_{ki})=-1}\rangle_{\mathcal{Z'}}}{\langle 
			e^{-\beta\mathcal{H}_{jk,ki}(S_{jk},S_{ki}=1)}+e^{-\beta\mathcal{
				H}_{jk,ki}(S_{jk}=1,S_{ki}=-1)}+e^{-\beta\mathcal{ 
				H}_{jk,ki}(S_{jk}=-1,S_{ki}=1)}+e^{-\beta\mathcal{ 
				H}_{jk,ki}(S_{jk},S_{ki})=-1}\rangle_{\mathcal{Z'}}},
		\end{aligned}
	\end{equation}
	the mean-field approximation yields the following relations for the 
	mean value of links and two-stars according to $Q$,
	\begin{equation}\label{eq9}
		p = \tanh(\beta Q) ,  \qquad 
		q(Q,J_{ijk},\beta) = \frac{e^{2\beta Q}-2\,e^{-2\beta J_{ijk} \tanh(\beta Q)} 
			+e^{-2\beta 
				Q}}{e^{2\beta Q}+2\,e^{-2\beta J_{ijk} \tanh(\beta Q)} +e^{-2\beta 
				Q}}.
	\end{equation}
	We can derive a self-consistent equation for $Q$. Let think of a network involves distinct groups of two-stars, each characterized by its number of members $N_{i}$ and weight $J_{i}$; then, we manage to write Eq. (\ref{eq6}) in the following form
	\begin{equation}\label{eq10}
	\begin{aligned}
	Q  &= \sum_{i} J_{i} \ q(Q,J_{ijk}=J_{i},\beta)  = 
	N_{1}J_{1}q_{1}+N_{2}J_{2}q_{2}+\cdots = (N-2)\sum_{i}P(J_{i})\ J_{i}\ 
	q_{i},
	\end{aligned}
	\end{equation}
	\end{widetext}
	where in the last line we have used $P(J_{i})$ the probability density of having two-stars with weight $J_{i}$. The factor $(N-2)$ indicates the total number of two-stars on a link and $q_{i}$ counts for the average of two-stars with weight $J_{i}$. From Eq. (\ref{eq9}) we see the average of two-stars, $q$, depends on the weight of the corresponding triad, $J_{ijk}$. 
	
	In the simplest case, we assume that there are two groups of two-stars 
	with the number of members $N_{1}$, $N_{2}$, and weights $J_{1}$, 
	$J_{2}$ on each link. Hence, Eq. (\ref{eq10}) is written as
	\begin{equation}
	\begin{aligned}
		Q &=N_{1}J_{1}q_{1} + N_{2}J_{2}q_{2}\\ 
		& = (N-2)\left\{P(J_{1})J_{1}q_{1} + P(J_{2})J_{2}q_{2}\right\},
	\end{aligned}
	\end{equation}
	$q_{1}, q_{2}$ indicate the average of two-stars with weights $J_{1} 
	, J_{2}$ respectively. Using Eq. (\ref{eq9}) to derive the mean value of 
	two-stars of each group 
	\begin{equation}
	\begin{split}
	&q_{1} = \frac{e^{2\beta Q}-2e^{-2\beta 
			J_{1}\tanh(\beta Q)}+e^{-2\beta Q}}{e^{2\beta Q}+2e^{-2\beta 
			J_{1}\tanh(\beta Q)}+e^{-2\beta Q}},\\
	&q_{2} = \frac{e^{2\beta Q}-2e^{-2\beta 
			J_{2}\tanh(\beta Q)}+e^{-2\beta Q}}{e^{2\beta Q}+2e^{-2\beta 
			J_{2}\tanh(\beta Q)}+e^{-2\beta Q}},\\
	\end{split}	
	\end{equation}
	multiplying $q_{1}$ in $N_{1}J_{1}$ and $q_{2}$ in $N_{2}J_{2}$ and summing them up, will come by a self-consistent equation based on $Q$
	\begin{equation}\label{eq13}
		Q = N_{1}J_{1}q_{1}+N_{2}J_{2}q_{2} \equiv f(Q;\;N_{1},N_{2},J_{1},J_{2},\beta),
	\end{equation}
	\begin{figure}[t]
		\centering
		\includegraphics[width=0.38
		\textwidth]{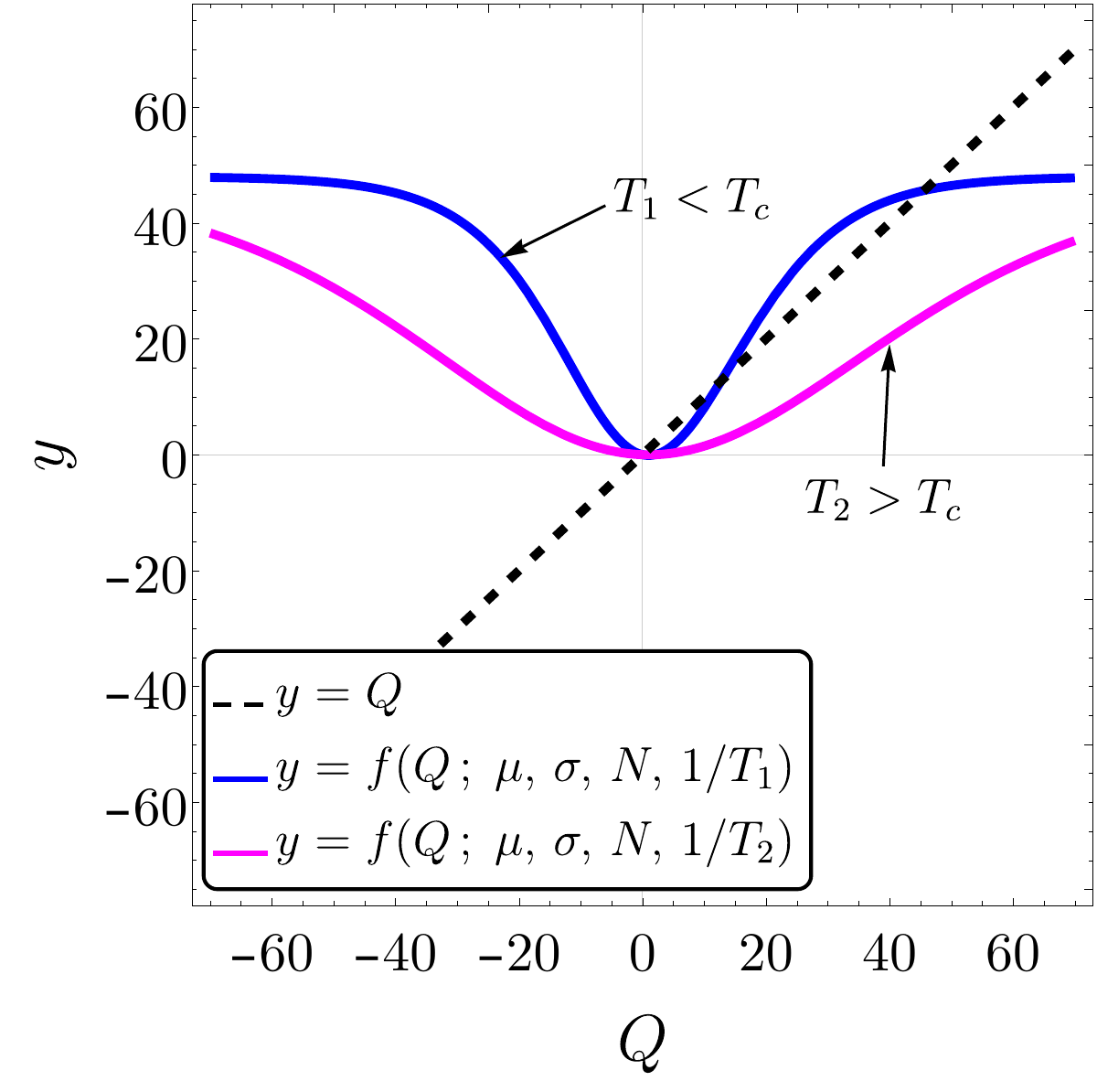}
		\caption{ Graphical analysis of Eq. (\ref{eq14}) for $N=50$, $T_{c}\simeq 28$.  For $T_{1}<T_{c}$ the equation has three fixed points which two of them are stable and one is unstable. For $T_{2}>T_{c}$ the equation has only one stable fixed point. }
		\label{fig:fig2}
	\end{figure}the intersection of the line $y=Q$ and the curve $y=f(Q;\;N_{1},N_{2},J_{1},J_{2},\beta)$ gives the solutions of Eq. (\ref{eq13}).

\begin{figure}
		\centering
		\includegraphics[scale = 0.3]{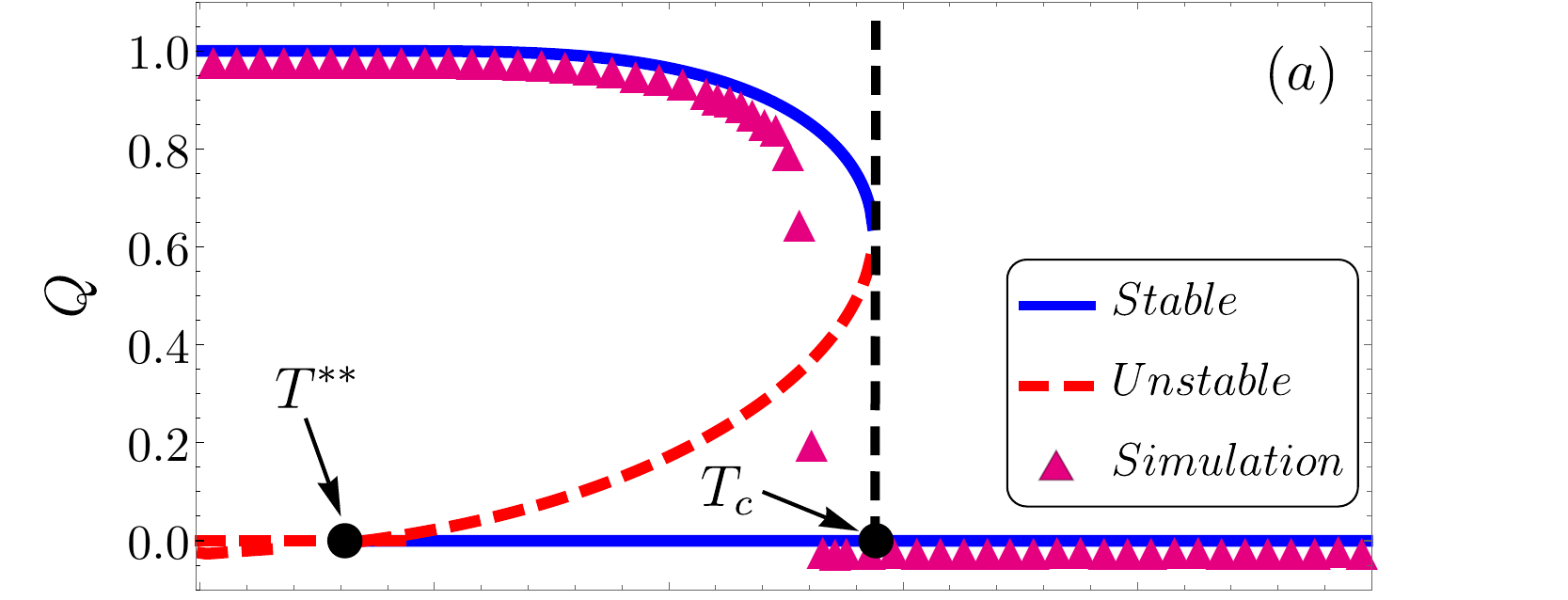}
		\includegraphics[scale = 0.3]{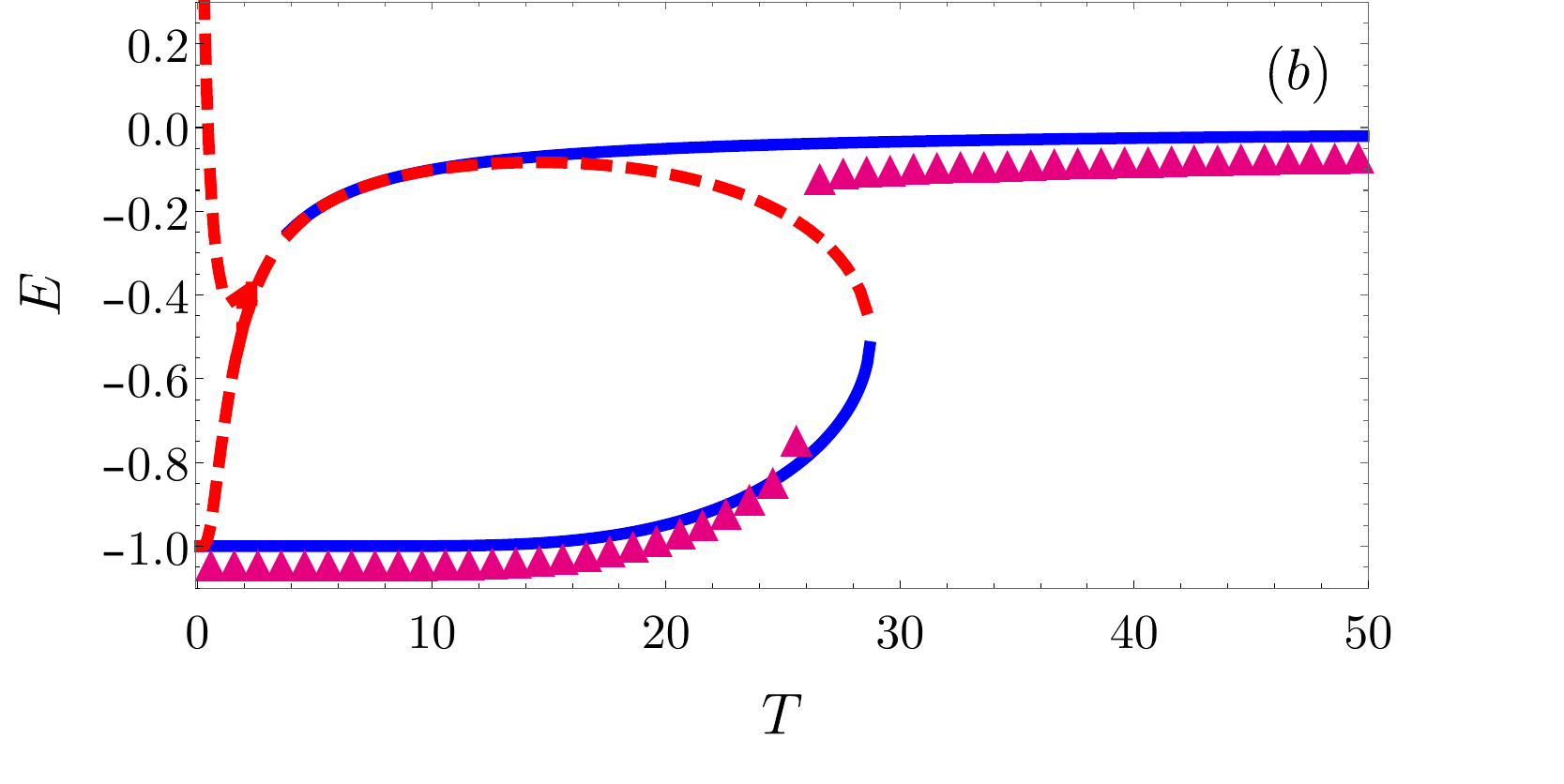}
		\caption{(a) Bifurcation diagram shows whether the normalized solutions of Eq. (\ref{eq14}), $Q^{\star}$, are stable or not. The diagram displays $T^{\star\star}$, the temperature that the zero changes from a stable to an unstable fixed point, and $T_{c}$, the critical temperature that indicates the phase transition of the system. The simulation result is also shown to confirm the accuracy of our analytical solution. (b) Bifurcation diagram of Eq. (\ref{eq17}) with normalized values and its corresponding simulation result.}
		\label{fig:fig3}
	\end{figure}

	Up to here, we succeed to derive a self-consistent equation by addressing Eq. (\ref{eq6}) in \textit{discrete manner} now let turn into \textit{continuous case} by converting the summation to integral as
	\small
	\begin{equation}\label{eq14}
		\begin{aligned}
			Q&=(N-2)\int_{-\infty}^{\infty}  J' \  P(J') \ q(Q,J',\beta) \ dJ'\\
			&=(N-2)\int_{-\infty}^{\infty} J' P(J')(\frac{e^{2\beta 
			Q}-2e^{-2\beta 
			J'\tanh(\beta Q)}+e^{-2\beta Q}}{e^{2\beta Q}+2e^{-2\beta 
			J'\tanh(\beta Q)}+e^{-2\beta Q}}) dJ' \\
			&\equiv f(Q;\;\mu,\sigma,N,\beta),
		\end{aligned}
	\end{equation}
	\normalsize
	this is the \textit{continuous} version of the self-consistent 
	equation which we solve numerically. (Fig.~\ref{fig:fig2}) shows a 
	plot of the forms $ y= Q$ and $y=f(Q;\;\mu,\sigma,N,\beta)$ as a 
	function of $Q$ for $N=50$, $\mu=1$, and $\sigma = 0.1$ in two 
	different temperatures. The intersection of line and curve gives the 
	solutions of Eq. (\ref{eq14}), which posses a first-order phase 
	transition between states of high and low temperatures. For $T>T_{c}$ 
	the system has only one stable fixed point $Q^\star=0$ which means 
	each link feels a zero effective field. This is the consequence of the 
	random distribution of two-stars of different signs and values on each 
	link. For $T<T_{c}$ the system has three fixed points, the middle 
	point is always unstable since the curve lies above $y=Q$ (the 
	derivative of the $f(Q;\;\mu,\sigma,N,\beta)$ respect to $Q$ is bigger 
	than one), the last point which stands for large $Q$ is always stable, 
	and the zero point that switches from a stable (solid line) to an 
	unstable (dashed line) fixed point in $T^{\star\star}$, 
	Fig.~\ref{fig:fig3}(a).

	The bifurcation diagram displays the kinds of possible solutions of 
	Eq. (\ref{eq14}) as a function of temperature 
	[Fig.~\ref{fig:fig3}(a)]. We see that by approaching the critical 
	point $T_{c}$ from above 
	two other fixed points will be created as well as $Q^{\star}=0$. The 
	critical temperature, $T_{c}$, indicates the phase transition of the 
	system and the difference in the number of solutions of Eq. 
	(\ref{eq14}). $T^{\star\star}$ is the temperature that the zero 
	switches from a stable to an unstable fixed point. The closeness of 
	the stable and unstable line in $T^{\star\star}$ point causes 
	disappearance of stable solutions for $T<T^{\star\star}$. The value of 
	both temperatures $T_{c}$, $T^{\star\star}$ depend on the variance.

	In the case of strong disorder (big variance), weights are so diverse 
	and values can be far away from the mean value of probability 
	distribution ($\mu=1$). Therefore, unlike the main idea of the 
	mean-field method, the environment of triads is not homogeneous and 
	each triad is experiencing different fields due to its surroundings. 
	It means we do not expect the mean-field method to works well in this 
	case. The remarkable effect of variance in the stability of the 
	network will be addressed in the proceeding sections where we compare 
	the simulation and analytical results. In the next section, we will 
	see in detail that the variance is playing a crucial role in either 
	achieving a reasonable analytical solution or its agreement with 
	simulation. Finally, we calculate the mean value of triads, 
	$r\equiv\langle S_{ij}S_{jk}S_{ki}\rangle$, by considering the 
	mean-field approximation (see the Appendix)
	\small	
	\begin{equation}
	\begin{aligned}
	& r(Q,J_{ijk},\beta) = \\
		&\frac{e^{3\beta Q+\beta J_{ijk}}-3e^{\beta Q-\beta 
				J_{ijk}}+3e^{-\beta Q+\beta J_{ijk}}-e^{-3\beta 
				Q-\beta J_{ijk}}}
		{e^{3\beta Q+\beta J_{ijk}}+3e^{\beta Q-\beta 
				J_{ijk}}+3e^{-\beta Q+\beta J_{ijk}}+e^{-3\beta 
				Q-\beta J_{ijk}}}.
	\end{aligned}
	\end{equation}
	\normalsize

	Based on the Hamiltonian Eq. (\ref{eq1}) the energy is the mean value 
	of weighted triads, $E = - 
	\langle\sum_{i>j>k}J_{ijk}S_{ij}S_{jk}S_{ki}\rangle$, which can be 
	written in the form of

	\begin{equation}
	E = -\sum_{i}P(J_{i})\ J_{i}\ r(Q,J_{ijk}=J_{i},\beta),
	\end{equation}
	 where $P(J_{i})$ is the probability density of finding a triad with 
	 weight $J_{i}$ and $r_{i}$ is the average of triads of weight 
	 $J_{i}$. Again considering \textit{continuous case} energy equation 
	 can be written in the following form:
	\begin{equation}\label{eq17}
	E  = -\int_{-\infty}^{\infty}  J' \ P(J') \ r(Q,J',\beta) \ dJ',
	\end{equation}
	 [Fig. \ref{fig:fig3}.(b)] pictures the analytical solution of energy 
	 for $\sigma=0.1$ and its simulation result. There is a reasonable 
	 agreement between simulation and analytical solutions.
	
	\subsection{Simulations}
	We start with a fully connected network. A link between two nodes $i$ 
	and $j$ has the values of $+1$ or $-1$ corresponds to friendly or 
	enmity relationship, respectively. We simulate a system with 50 nodes, 
	$\mu = 1$ and two different values of variance $0.1$, $10$. We 
	thermalize our system with a given temperature by the Monte Carlo 
	method. We are using the Metropolis algorithm; in each iteration, for 
	a given temperature a link is selected randomly, and based on our 
	Hamiltonian we calculate the energy difference $\Delta E = E_{2}-E_{1} 
	$  if the energy of configuration after flip $E_{2}$ is less than 
	prior $E_{1}$ ($\Delta E <0$), the selected link is flipped otherwise 
	flip will be accepted with a probability equal to ``exp$(-\beta \Delta 
	E)$" where ``$\beta$" is equal to the inverse of the temperature. 
	Since we consider triads with different weights, it 
	matters from which triad the random link is. Flipping a link that 
	belongs to a big weighted triad will change the energy of the system 
	significantly. Here, the role of the variance would appear. In weak 
	disorder a large percentage of weights are around the mean value of 
	probability distribution ($\mu =1$), so triads have equal weights 
	approximately, whereas in strong disorder regime, 
	weights are diverse and competition between triads of various weights 
	prevents the network to reach a balanced state.
	\subsubsection{Weak disorder} 
	In the case of small variance, $\sigma=0.1$, the majority of weights 
	are positive (mean of probability was set to be in positive values 
	$\mu=1$). In high temperatures, fluctuations are high and the total 
	energy of the system is zero. As temperature decreases 
	the links switch and triads change into balance whether they are 
	positively or negatively weighted. Thermal fluctuation can overcome 
	small weighted triads and flip links to satisfy energy minimization. 
	Although triads of different weights compete with each other to 
	determine the network destination, positive triads overcome the 
	contest since they are the majority in terms of numbers and the 
	balanced state of the network occurs as triads are in balanced 
	configurations in the definition of $J>0$. The same will happen if the 
	mean of probability was set to be in negative e.g, $\mu=-1$, with the 
	difference that negative triads win the competition and a balanced 
	state occurs in the definition of $J<0$.
	
\begin{figure}[t]
		\centering
		\includegraphics[scale = 0.416]{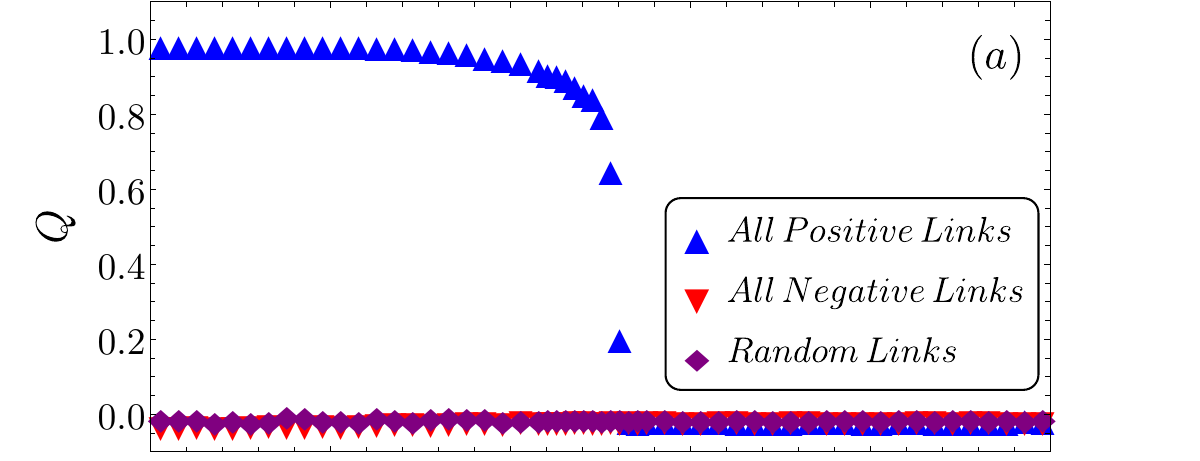}		
		\includegraphics[scale = 0.3]{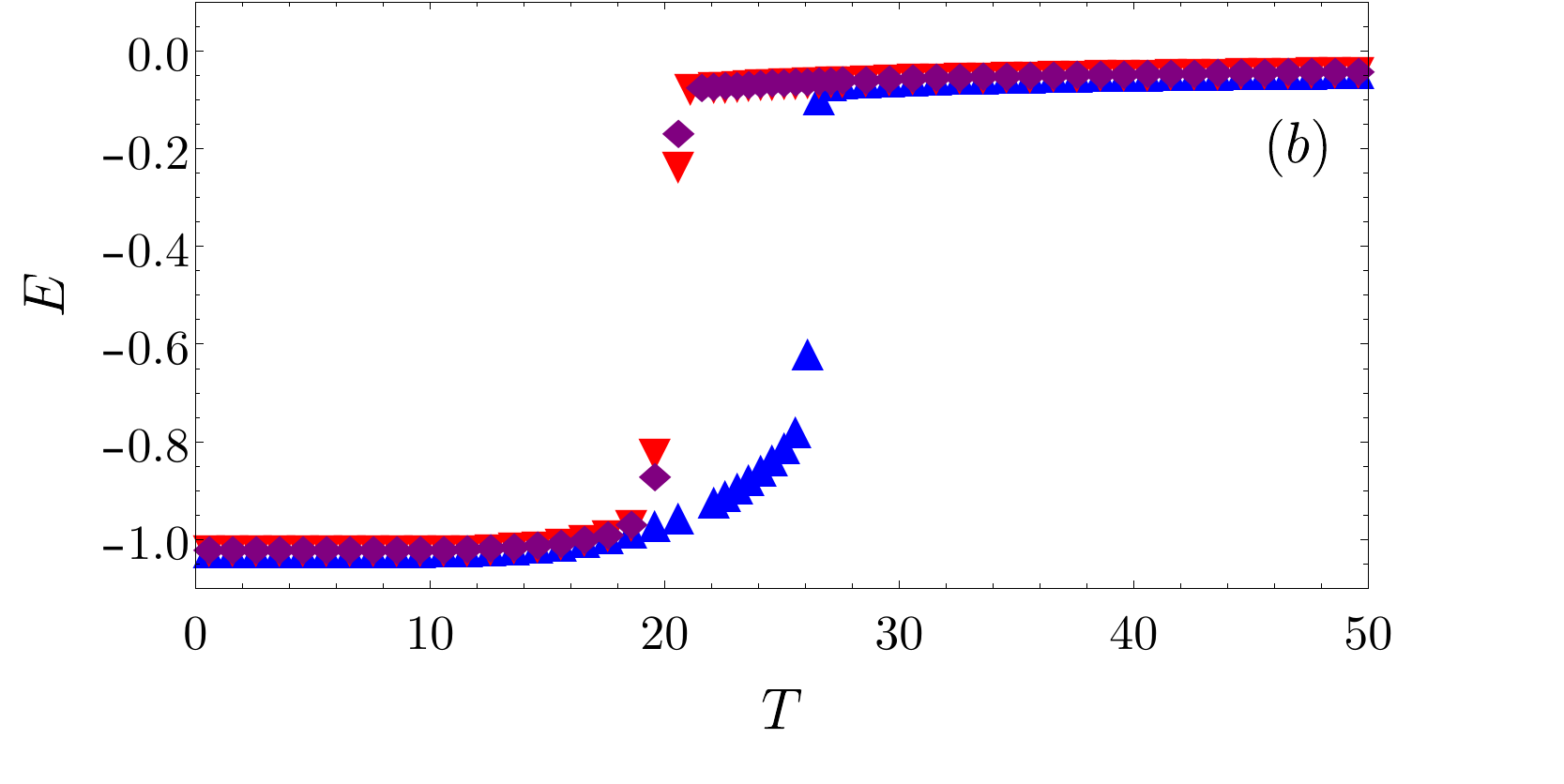}
		\caption{(a) The effective field (the mean value of weighted 
		two-stars), $Q$, vs. temperature. In high temperatures, each link 
		feels a zero field because we have equal number of positive and 
		negative two-stars. The final state of positive initial condition 
		in low temperature is heaven so, $Q=1$. For other initial 
		conditions, the final state is bipolar and an equal ratio of 
		positive and negative effective field results in $Q=0$. (b) The 
		energy of network vs. temperature for different initial states. In 
		high temperature energy is zero since the system is in the random 
		state and by decreasing temperature system is led to the balanced 
		state.}
		\label{fig:fig4}
	\end{figure}

	Figure.~\ref{fig:fig4}(a) displays the effective field each link feels 
	respect to the temperature. The balanced final state of the system 
	depends on the initial condition. In the case of all positive links, 
	the final state is heaven so the normalized effective field in low 
	temperatures is equal to one. For other initial conditions, the final 
	state is bipolar and each link feels approximately equal pressure from 
	positive and negative fields so the normalized effective field equals 
	zero. Figure.~\ref{fig:fig4}(b) shows the energy simulation results 
	versus temperature for different initial states (all links positive, 
	all links negative, random links). The transition between balanced and 
	random states happens in two different temperatures for all positive 
	initial condition and two other conditions. Starting 
	simulation with an all links positive network and decreasing the 
	temperature, the system reaches to the balanced state 
	at higher temperature. We see that the minimum energy 
	of the network occurs in the mean value of the probability 
	distribution which is in agreement with our expectation since the 
	network mostly consists of triads with weights around $\mu=1$. 
	Increasing the variance will increase the ratio of triads of variously 
	weighted. We will address the big variances in the proceeding section.
	\subsubsection{Strong disorder}
	 For strong disorder, the number of triads of different weights 
	 (positive and negative) are comparable to each other, e.g., for 
	 $\sigma>4$ the percentage of negative triads is more than $40$ 
	 therefore, there is a competition between negatively weighted triads 
	 and positive ones. There is no end to this competition even if we 
	 increase Monte-Carlo steps by order of magnitude; hence, we see that 
	 the system does not reach the global minimum at low temperatures. 
	 Generating random weights from a Gaussian distribution with a big 
	 variance will result in values that are not necessarily around the 
	 mean. Here, the diversity of weights and the 
	 comparable number of positive and negative triads cause the network 
	 to be unable to reach a balanced state.  Notice we do not have such a 
	 situation in the weak disorder case since almost all triads have 
	 roughly equal weights.
	 
 	\begin{figure}[t]
 	\centering
	\includegraphics[scale = 0.416]{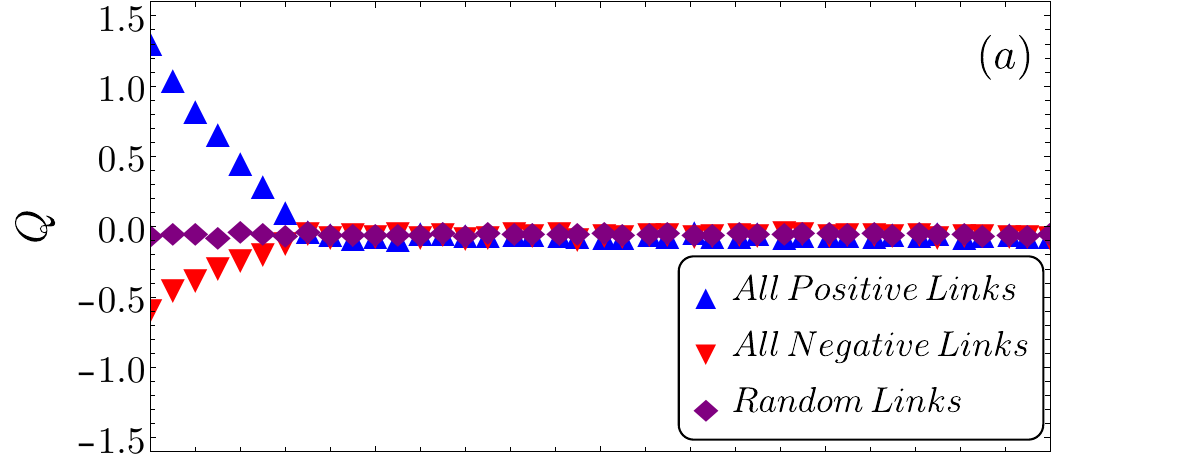} 	
	\includegraphics[scale = 0.3]{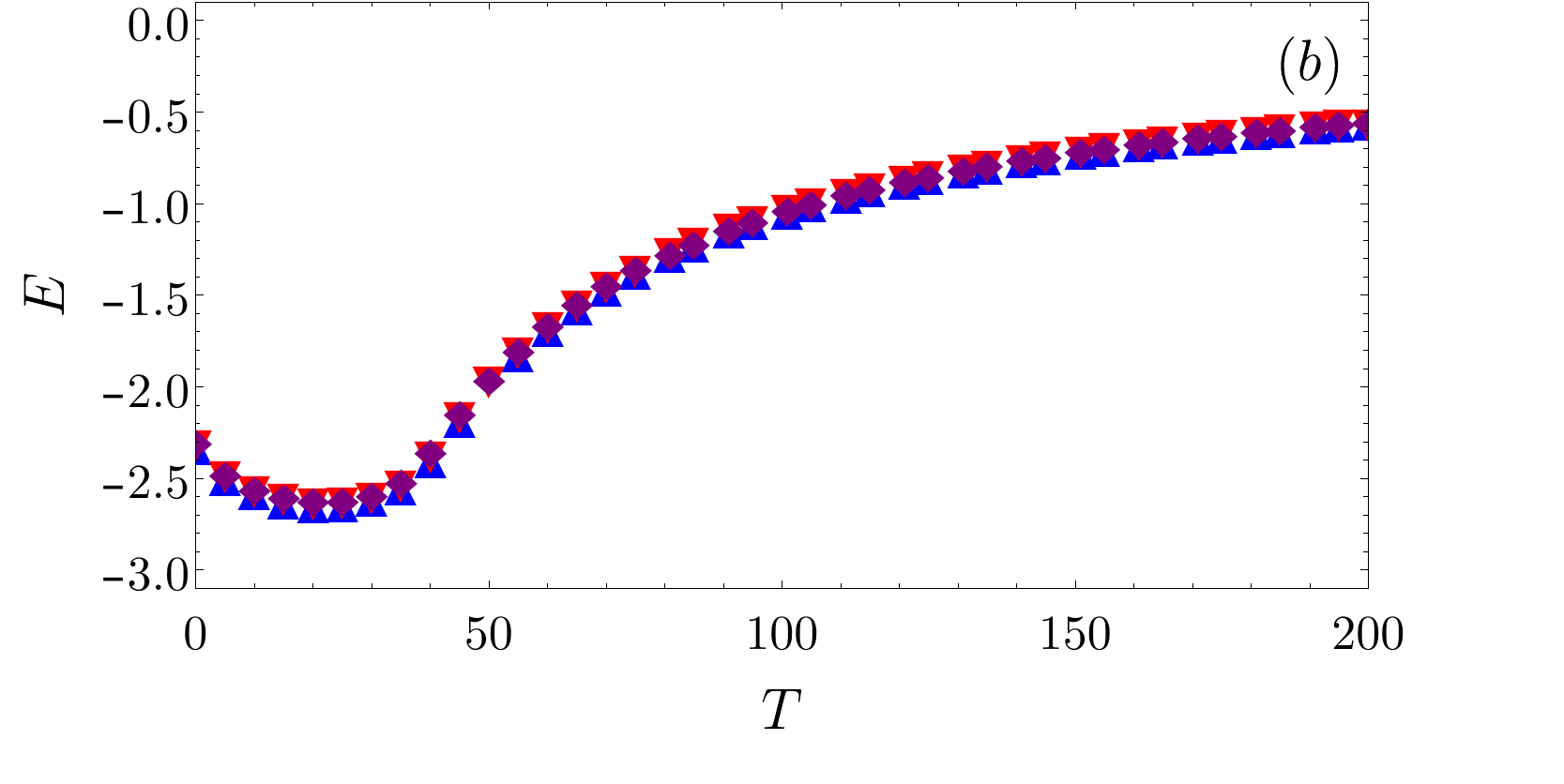}
	\caption{(a) The effective field, Q, vs. temperature. In high 
	temperatures, each link feels a zero field due to the random 
	distribution of two-stars. Because of the diversity of weights in big 
	variance, the behavior of this quantity deviates from the weak 
	disorder case in low temperatures. (b) The energy of the network vs. 
	temperature. The interaction between triads with various weights 
	causes zero energy in high temperatures. In low temperatures, the 
	existence of unbalanced triads causes energy increases so, the system 
	does not reach a balanced state .}
	\label{fig:fig5}
	\end{figure}
\begin{figure}[ht]
		\centering		
		\includegraphics[scale = 0.25]{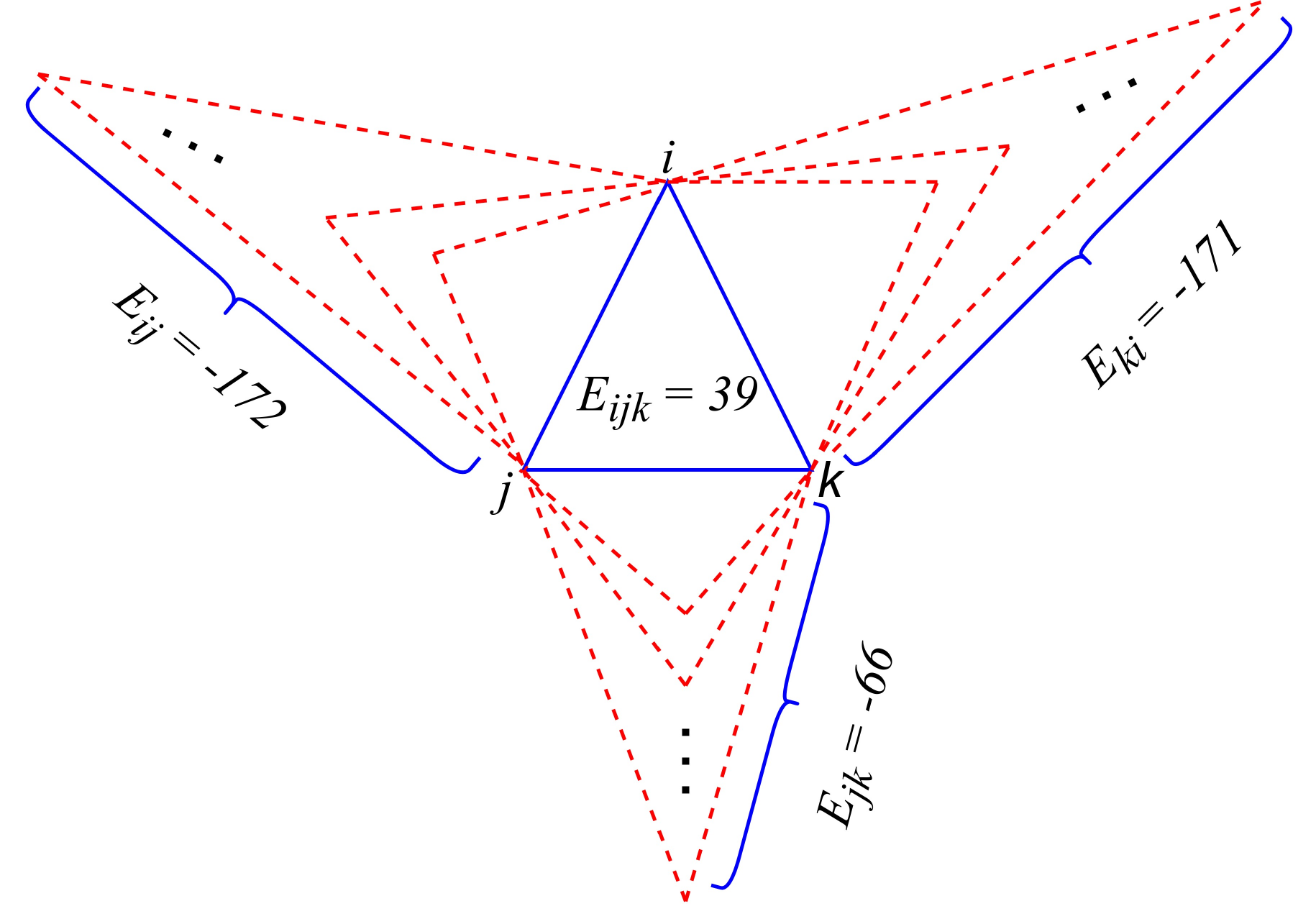}
		\caption{A sample unbalanced triad, $E_{ijk}=39$, 
		is depicted. Each link shares in several triads (here, we have 
		shown only three of them). The total energy of triangles on each 
		link is shown, e.g., $E_{ij}$ is the sum of the energy of 
		triangles on the link $i,j$. The negative value of energy on each 
		link shows that the majority of triads were balanced.}
	    \label{fig:fig6}
	\end{figure}

	Figure.~\ref{fig:fig5}(a) displays the effective field each link 
	feels, depends on the initial condition. The
	simulation for strong disorder shows in high 
	temperature, $T \gg T_{c} $, the random distribution of positive and 
	negative two-stars results in zero effective field on each link and in 
	$ T \ll T_{c} $, the existence of disorder in the network results in a 
	deviation in the behavior of effective field from the weak disorder 
	regime, Fig. 4(a). Beginning with the positive initial condition, in 
	small disorder case the network reaches the heaven state and all 
	triads are of the kind $\{+,+,+\}$, likewise, in big variance case, it 
	is more probable to have balanced triads of a kind $\{+,+,+\}$, 
	$\{-,-,-\}$ for $J>0$ and $J<0$ respectively so the number of positive 
	two-stars overcomes. For the case of the negative initial condition, 
	in the small disorder case, the network has an equal number of 
	positive and negative two-stars since it achieves bipolar state and 
	the balanced triads of kind $\{-,-,+\}$, similarly in strong disorder 
	case the balanced triads are more probable to be of a kind 
	$\{-,-,+\}$, $\{+,+,-\}$ for $J>0$ and $J<0$ respectively. In these 
	mentioned kinds of triads two-stars are averagely negative so we see a 
	negative value of $Q$ in $T \ll T_{c}$.
	Fig.~\ref{fig:fig5}(b) indicates the dynamics of the 
	network with respect to the temperature for 
	$\sigma=10$. It shows, all initial states diagrams overlay, in 
	contrast with the small variance that 
	positive initial state is isolated from two others 
	Fig.~\ref{fig:fig4}(b). Our horizon toward the system’s evolution in 
	the strong disorder regime will be opened by Fig.~\ref{fig:fig6}.

	It represents, the only possible situation that the unbalanced triads 
	can survive in the network is to be surrounded by a large number of 
	balanced ones. The general strategy is to update the links in each 
	temperature. If switching a link results in by a negative energy 
	difference $\Delta E <0 $, then the flip will be accepted; otherwise, 
	the condition of a link to be flipped is measured by Boltzmann 
	probability 
	$ p \propto$ exp$(-\Delta E/T)$ and the temperature gives a chance for 
	the links to be flipped. Investigating the structure of the network in 
	three regimes of temperature yields:
	
	(i) $T \ll T_{c}$, it is more probable that the system getting stuck 
	on the local minimum and links do not have the chance of flip, hence, 
	the system can not escape from the local minimum and unbalanced triads 
	froze in their states. In the Fig. 6, we chose a sample 
	unbalanced triad and calculated the sum of the energy of triads on 
	each edge. The flipping of each link results in a growth in energy 
	$(\Delta E > 0)$ and the Boltzmann probability of flipping is about to 
	zero $ p \cong 0 $ so unbalanced triads would not change. Then the 
	network has more unbalanced triads in these temperatures and we see 
	that the energy increases.
	
	(ii) $T \simeq T_{c} \simeq 28$, with increasing temperature, links 
	have 
	the chance of flipping due to the Boltzmann probability and some of 
	the unbalanced triads change into balanced. The energy minimum occurs 
	in temperatures around $20- 30$ where the number of unbalanced triads 
	is 
	the least with respect to the other temperatures.
	
	(iii) $T \gg T_{c}$, when temperature tends to infinity, based on the 
	Boltzmann probability, links have the chance of flipping even if the 
	energy increases. The temperature provides the possibility that 
	unbalanced triads exist in the network. Therefore, the network is a 
	combination of balanced and unbalanced triads which results in zero 
	energy.
	
	 From another point of view, the diagram of the energy 
	 versus temperature in the strong disorder regime, 
	 [Fig.~\ref{fig:fig5}(b)] shows, in the case of big variance, energy 
	 increases at low temperature, although we expect to see the reduction 
	 in the number of unbalanced triads and the minimization of energy as 
	 temperature decreases. This figure is schematically similar to the 
	 resistivity versus temperature diagram in the Kondo effect, 
	 \cite{kondo1,kondo2}; we expect to see low resistivity in low 
	 temperatures, but the existence of impurity in the lattice which 
	 disturbs the periodic layout of the lattice results in an unusual 
	 behavior of some metals. We emphasize that we have no claim on the 
	 Kondo effect concept in this study; merely there is a similar 
	 behavior between these two phenomena. As we mentioned above the 
	 mean-field approximation does not work well in this regime due to 
	 heterogeneity of weights. Here the difference between our proposed 
	 model and structural balance theory emerges. We see the energy will 
	 reach zero in temperatures of the order $10^{4}$. 
	\section{CONCLUSIONS}   
	Here, we proposed a model that takes into account weights in the 
	structural balance theory. Results show that it is not a simple 
	modification to unweighted balance theory and our analysis 
	demonstrates the importance of variance in achieving a balanced state. 
	We studied the model theoretically through the mean-field method that 
	suites perfectly the Monte Carlo simulation results in the weak 
	disorder regime. Looking at the dynamics of the system in a completely 
	connected network under two regimes revealed that the system posses a 
	first-order phase transition because of a discontinuous jump in the 
	energy of the system. Since increasing the variance corresponds to the 
	increase in the number of negative triads or more disorder in the 
	network, then it takes longer steps for the system to reaches a 
	balanced state in low temperatures in contrast to small variances 
	hence, the critical temperature depends on the variance. 
	In the process of the evolution, although each kind 
	of triads whether positive or negative wants to make the 
	network balanced in its definition, we see the unbalanced triads still 
	persist in low temperatures 
	in strong disorder, unlike weak disorder and balance 
	theory. This result represents a pseudo-Kondo effect behavior in the 
	energy changes versus temperature which states the remarkable 
	difference between our model and balance theory. Our proposed model 
	goes beyond the equally weighted triads and pays attention to 
	differences of relations so it can serve as a practical tool in 
	dealing with real weighted networks.
	 
	\begin{acknowledgments}                
	We appreciate  S.  Oghbaiee,  S.  Cheraghchi for fruitful discussions. We gratefully acknowledge the Center of Excellence in Cognitive Neuropsychology.
	\end{acknowledgments} 
               
	\appendix
	\begin{appendices}
		\section{Calculation of The Mean Values Triads}\label{appendix}
		\label{appendix:mean quantity}
		In this part, we want to calculate the mean value of triads. We write the Hamiltonian as $H = H'+ H_{\Delta}$, where $ H_{\Delta}$
		\begin{equation}
		\begin{aligned}
		H_{\Delta}& = -S_{ij}(\sum_{l\neq{i,j,k}} 
		J_{ijl}S_{il}S_{lj})-S_{jk}(\sum_{l\neq{i,j,k}}J_{jkl}S_{kl}S_{lj})\\
		&-S_{ki}(\sum_{l\neq{i,j,k}}J_{ikl}S_{il}S_{lk})-J_{ijk}S_{ij}S_{jk}S_{ki} ,
		\end{aligned}
		\end{equation}
		consist of terms dedicates to each link $S_{ij}$ ,  $S_{jk}$ and $S_{ki}$ individually and the last term denotes a triad of all those. 
		We have
		\begin{equation}
		\langle{S_{ij}S_{jk}S_{ki}}\rangle=
		\sum_{\{S_{ij},S_{jk},S_{ki}\pm1\}} P(S_{ij},S_{jk},S_{ki})S_{ij}S_{jk}S_{ki} .\\
		\end{equation}
	    Keep all the calculations similar to what has been done for two-stars, again in the mean-field approximation we estimate each of three inequality summations by two inequality, for convenient use of $Q$ definition. Considering all possible configurations of triplet $S_{ij}S_{jk}S_{ki}$, there will be eight different forms, where $ H^{abc}_{\Delta}\equiv\ H_{\Delta}(S_{ij}=a,S_{jk}=b, S_{ki}=c) $, and we write it as
		\begin{equation}
		\begin{aligned}
		-\beta H^{+++}_{\Delta}& =  3\beta Q+\beta J_{ijk},\\
		-\beta H^{---}_{\Delta}& = -3\beta Q-\beta J_{ijk},\\
		-\beta H^{++-}_{\Delta}& = \beta Q-\beta J_{ijk},\\
		-\beta H^{+-+}_{\Delta}& = \beta Q-\beta J_{ijk},\\
		-\beta H^{-++}_{\Delta}& = \beta Q-\beta J_{ijk},\\
		-\beta H^{--+}_{\Delta}& = -\beta Q+\beta J_{ijk},\\
		-\beta H^{-+-}_{\Delta}& = -\beta Q+\beta J_{ijk},\\
		-\beta H^{+--}_{\Delta}& = -\beta Q+\beta J_{ijk},
		\end{aligned}
		\end{equation}
		\begin{widetext}
			\begin{equation}\label{opensquares}
				r(Q,J_{ijk},\beta) = 
			\frac{e^{3\beta Q+\beta J_{ijk}}-3e^{\beta Q-\beta 
					J_{ijk}}+3e^{-\beta Q+\beta J_{ijk}}-e^{-3\beta 
					Q-\beta J_{ijk}}}
			{e^{3\beta Q+\beta J_{ijk}}+3e^{\beta Q-\beta 
					J_{ijk}}+3e^{-\beta Q+\beta J_{ijk}}+e^{-3\beta 
					Q-\beta J_{ijk}}}.
			\end{equation}
		\end{widetext}
	\end{appendices}

\end{document}